# Sol-Gel and Gel-Sol Transitions of Biodegradable *Chlorella-k-Carrageenan* Composites; Photon Transmission Study

İrem Doğruoğlu, Yağmur Akarsu, Çağla Selen Şen, Prof. Önder Pekcan, Assoc. Prof. Bengü Özuğur Uysal

**ABSTRACT:**

Microalgae can be used in the packaging sector due to their ability to synthesize bioactive compounds that are favorable in the production of biodegradable packaging. In this work, to create an alternative to plastic packaging, we studied microalgae *Chlorella* in the presence of *k-Carrageenan* to form biodegradable packaging using various approaches such as gel formation, film creation, and photometric analyses. Here we have measured Gel-Sol and Sol-Gel energies by employing Arrhenius Law on photon transmission curves, produced during these hysteresis-type phase transitions. The combination of *Chlorella, k-Carrageenan,* NaCl, and water could form a gel that had a high biodegradation rate with no photodegradation obtained from the experiments. At the same time, our findings indicated that the material obtained from these processes could be a considerable option that degrades much faster than the plastic.

## 1. INTRODUCTION

Microalgae have recently become a point of interest worldwide because of their vast application potential in many areas. They can be used as biofuels, food sources, and bioactive medicines, as well as being renewable and sustainable. They are the perfect replacement for regular liquid fossil fuels regarding cost, sustainability, and environmental issues. They can also turn atmospheric $CO_2$ into other bioactive metabolites. Several kinds of microalgae have been researched for their potential, such as *Carrageenan* and *Chlorella* [1].

*Carrageenan* is the generic name of a sulfated polysaccharide, which is extracted from red seaweeds from the class Rhodophyceae and is heavy in molecular weight. It consists of galactose and anhydrogalactose, connected via glycosidic bonds. It is preferred in the food industry for its ability to gelate, thicken, and emulsify. It is also used in some other commercial products, as well as pharmaceutical formulas as it has shown potential as being anticoagulant, anticancer, antihyperlipidemic, and immunomodulatory [2].

*Chlorella Vulgaris*, belonging to the genus *Chlorella*, is one of the most noteworthy green microalgae. It can mainly be found in various aqueous environments. It has a high capability for photosynthesis and is known for its fast growth under various conditions. *Chlorella vulgaris* is mainly made of proteins, carbohydrates, lipids, pigments, vitamins, and minerals. Most of *Chlorella Vulgaris'* dry body weight is protein content, around 43-58%, and is required for basic functions like the growth of the cell [3].

There are many areas in which microalgae are employed. One of the most common areas is the food industry, thanks to its nutritional capacity. They can also be used to create pigments. In addition, microalgae are promising in medicine, as they can be used for sourcing bioactive compounds beneficial to health, such as compounds with anti-atherosclerotic or anti-cancer properties [3]. Due to their high lipid content, along with their high biomass productivity, microalgae can be a new source of biofuel. When compared with chemical fertilizers, microalgae can be used to replace them as bio-stimulants. As well as the previous areas mentioned, microalgae are great sources to produce bioplastics. Bioplastics obtained from microalgae can be used in food packaging, medical hygiene, and food transportation. Bioplastics are favored over petrochemical polymers due to their low toxicity, barrier properties, and biodegradability [4].

Industries like the packaging sector have always placed a strong mindset on maximizing product protection and package durability without considering enough to the disposability of materials [5]. To overcome issues that may arise from disposability, several methods like biodegradable products have been created. Substances or materials that may quickly break down by bacteria or any other natural creatures without contributing to pollution are categorized as biodegradable [6].

Biodeterioration, biofragmentation, and assimilation are the three phases of the biodegradation process. The term biodeterioration alters the material's mechanical, physical, and chemical characteristics. This stage takes place when the substance encounters abiotic elements to weaken the material's structure allowing further breakdown. Secondly, a polymer is biofragmented during the lytic phase of bond cleavage, which produces oligomers and monomers in its substitute. Lastly, the products of biofragmentation will proceed to the assimilation step [7].

In connection with a substance's capacity for persistence and bioaccumulation in the ecosystem, photodegradation is also a crucial factor in determining a substance's environmental outcome. A substance's capacity for degradation of subjection to light that is typically ultraviolet (UV) radiation from the sun is referred to as photodegradability [8]. The substrate concentration, the quantity of photocatalyst, the pH of the solution, the overall temperature, the time and intensity of illumination, the surface area of the photocatalyst, the dissolution of oxygen in the reaction medium, the structure, and the nature of the photocatalyst and substrate can affect the rate of photodegradation [9]. The absorbance spectra from a previous study conducted by Pfendler et. al. was taken as a reference to examine the UV results [10].

Many processes, such as the rupturing of chemical bonds and the production of free radicals, can lead to photodegradation. A substance can absorb photons with light energy when it is illuminated by elevating the substance to a higher energy level. When moving higher and more reactive energy states occur, that substance may undergo reactions like a chemical bond breakdown. Once the creation is done, free radicals can trigger chain reactions in which they interact with other molecules to create more free radicals. Till the free radicals are neutralized or the reaction is stopped, the photodegradation process can continue. [11].

In this work, the main purpose was to create a microalgae-based alternative to plastic packaging. Photometric analyses were used to examine the gel formation mechanisms, with a focus on transmittance, absorbance, and Arrhenius plots. In addition to that, heating (gel-sol) and cooling (sol-gel) were performed to examine the materials during the phase transitions. Lastly, biodegradability and photodegradability tests were done on the optimal sample obtained through the photometric analyses.

## 2. METHODOLOGY

### 2.1 Gel Formation

Gel formation can be described as a spontaneous reaction from a basic polymer dispersion, particle suspension, and externally adjustable circumstances of temperature or solution composition [12]. During gel formation, a sol-to-gel transition occurs. This transition consists of the aggregation of particles, which eventually leads to a network spanning the entirety of the volume of the container [13]. Gel formation can be induced by either physical factors, such as heat and pressure, or chemical factors, such as ion addition [12].

Heat-induced gel formation is one of the most used gel formation methods. It consists of two steps. The first one is the unfolding or dissociation of the molecules caused by the input of energy in the beginning. The second one is the association and aggregation of unfolded molecules, which establishes high molecular weight complexes [14]. To achieve heat-induced gel formation, the aqueous mixture containing the gelator is simply heated up, followed by cooling [15].

## 2.2 Film Formation

Film formation is the process of forming a thin film of a film-forming substance. Usually, this substance is in liquid form and applied on a solid substrate to obtain a film. The application is usually done by spraying the polymer on the solid substrate, which can be done after the polymer is dissolved in a proper solvent. After being sprayed on the substrate, the droplets spread and after the solvent is evaporated, a film is formed, as shown in Figure 1 [16].

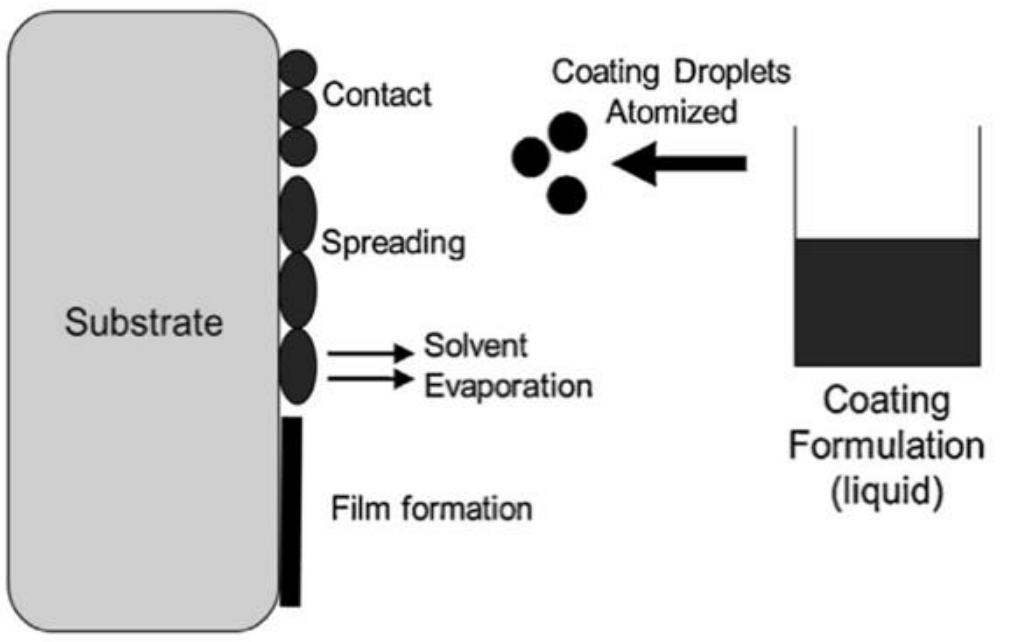

Figure 1. Working principle of film formation [16].

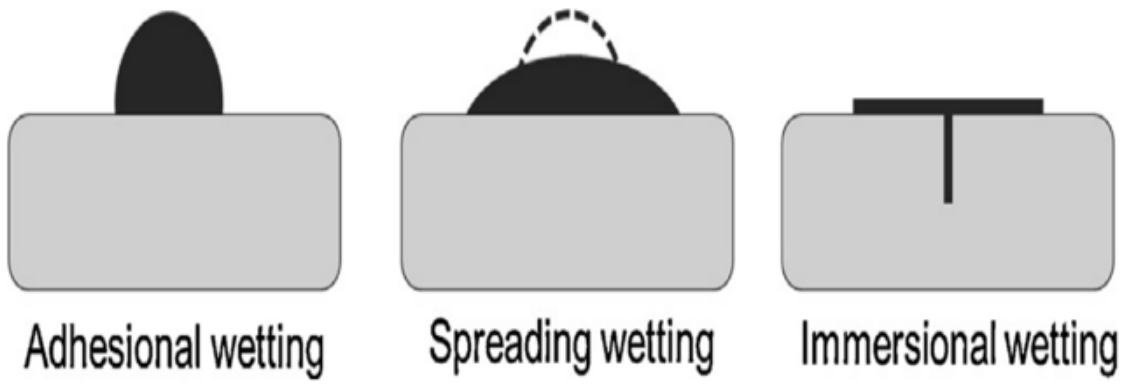

Figure 2. Adhesional, spreading and immersional wetting [16].

Solvent evaporation occurs during film formation with polymer solutions, enabling polymer chains to interpenetrate and create a gel phase. The rate of solvent evaporation is critical for film formation since too slow or too fast evaporation might result in overheating in substrate or dry polymer droplets. The rate of evaporation can be affected by factors such as humidity, temperature, air velocity, and atmospheric pressure. The film formation also depends on the droplet's ability to wet the surface, which is influenced by both the solid substrate and the solution. There are three types of wetting processes, displayed in Figure 2: Immersional, where no change in the air-liquid interface occurs; spreading, where there's an air-liquid interface; and adhesional, where there is no air-liquid interface [16]. The rate of wetting depends on factors such as substrate and coating formulation, along with processing conditions [17].

## 2.3 Photon Transmission Technique

By measuring the reactivity and light absorbance of gels concerning their chemical characteristics at a certain wavelength, a photometric analysis is a technique that can be employed to assess the creation and stabilization of gels. Absorbance (A) can be described as the light that a solution can absorb. On the other side, the transmittance

(T) is the solution's capacity to let the light pass through [18]. After the gel formation process, photometric analyses are performed to observe the solvent-to-gel transition and the reversible gel formation process transition in depth [19].

A spectrophotometer, a tool for determining the absorbance or transmission of light of a substance, can be used to detect gel formation. The existence of the gel network in a gel-forming solution affects the amount of light scattering. Hence, by monitoring changes in the sample's light scattering characteristics, the gel formation procedure can be observed. The concentration of gel-forming chemicals, the pH, and the temperature can all be optimized by utilizing photometric measurements [20]. By altering optical parameters, light transmission can be changed in a gel, allowing for the measurement of gel formation dynamics and overall strength [21]. The absorbance versus wavelength graph from a study by Yatirajula et. al. was taken as supplementary data [22].

## 3. MATERIALS

### 3.1 *Chlorella*

*Chlorella* and distilled water were used initially. All the samples were mixed in 5 mL beakers, obtained from the brand ISOLAB. The *Chlorella* used was in powder form, and was obtained commercially, from the brand Kuru Yeşil. The packaging and the energy and nutrients in the powdered *Chlorella* are shown below in Table 1. The *Chlorella* used in each sample was measured by using Series 390 Semi-Micro analytical balance, by the brand Precisa. At the same time, different samples that contained *Chlorella*, distilled water, and acetone obtained from Sigma-Aldrich were prepared. To properly stir each sample, magnetic stirrer bars obtained from Fischer Scientific were used. To obtain the gel form, the samples were heated and stirred using an MTOPS MS300HS Hot Plate & Stirrer.

**Table 1.** Energy and nutrients table for *Chlorella*.

| Energy and Nutrients | For 100 grams |
|---|---|
| **Energy** | 360 (kcal) |
| **Total Fat** | 1.2 (g) |
| **Saturated Fat** | - |
| **Trans Fat** | - |
| **Carbohydrate** | 36 (g) |
| **Sugars** | 4 (g) |
| **Fiber** | 0.4 (g) |
| **Protein** | 64 (g) |
| **Salt** | - |

The materials used to achieve the gel formation are *Chlorella*, *k-Carrageenan*, and sodium chloride.

## 4. EXPERIMENTAL

### 4.1 Sol-Gel/Gel-Sol Phases Observation with UV-VIS Spectrophotometer

For this part of the experiment, the best samples from each set of samples were poured into quartz cuvettes from the brand Sigma-Aldrich and placed into a 752N Plus UV-VIS Spectrophotometer from the brand HINOTEK. For the heating part, a hot water bath was prepared in a 1 L beaker from the brand ISOLAB, in which water was poured into the beaker and heated on top of the MTOPS MS300HS Hot Plate & Stirrer. A PeakTech 2010 model digital multimeter was placed inside quartz cuvettes to measure the temperature.

One Uni-T brand UTP3315TFL-II model regulated DC power supply and one PHYWE brand Stelltrafo model power supply were used to power a smaller 12V motor source, which was used to transfer the water around the entire setup. This setup was connected by using pipes and a copper coil. For the cooling part, the same setup was used, by removing the copper coil and then removing the beaker from the hot plate.

### 4.2 Biodegradability Analyses

For this test, the sample with the best gel formation was obtained in film form and buried in soil obtained from school grounds. Four 25 mL beakers from the brand ISOLAB were used to bury four different amounts of samples in soil. To check if a bigger sample could dissolve in soil, a fifth sample was prepared as a film by filling half of a circular ice cube mold. To test for the water resistance, dry components were measured by using a Series 390 Semi-Micro analytical balance, by the brand Precisa. Next, a 50 mL beaker obtained from the brand Boru Cam, and another 25 mL beaker obtained from the brand ISOLAB were used to create a mold in the shape of a cup.

## 5. RESULTS

### 5.1 *Chlorella* and *Carrageenan*

Five different samples were prepared in separate 5mL beakers to get a detailed observation. Detailed amounts of components in samples were prepared and given in Table 2.

**Table 2.** Detailed amounts of components in samples.

| Sample # | *Chlorella* *(g)* | *k-Carrageenan* *(g)* | *NaCl* *(g)* | *Distilled water* *(mL)* |
|---|---|---|---|---|
| **Sample 1** | 0.06 | 0.06 | 0.3 | 5 |
| **Sample 2** | 0.06 | 0.12 | 0.3 | 5 |
| **Sample 3** | 0.06 | 0.18 | 0.3 | 5 |
| **Sample 4** | 0.06 | 0.24 | 0.3 | 5 |
| **Sample 5** | 0.06 | 0.30 | 0.3 | 5 |

When sample preparations were done using an analytic balance and magnetic stirrers bar, all samples were heated from 20 to 80 C° for around 1 hour at room temperature (20 C°). In the end, samples were cooled for around 5 minutes at room temperature, and proper gel formation was obtained at the highest level (Figure 3). All five samples were poured into separate quartz cuvettes to perform photometric analyses.

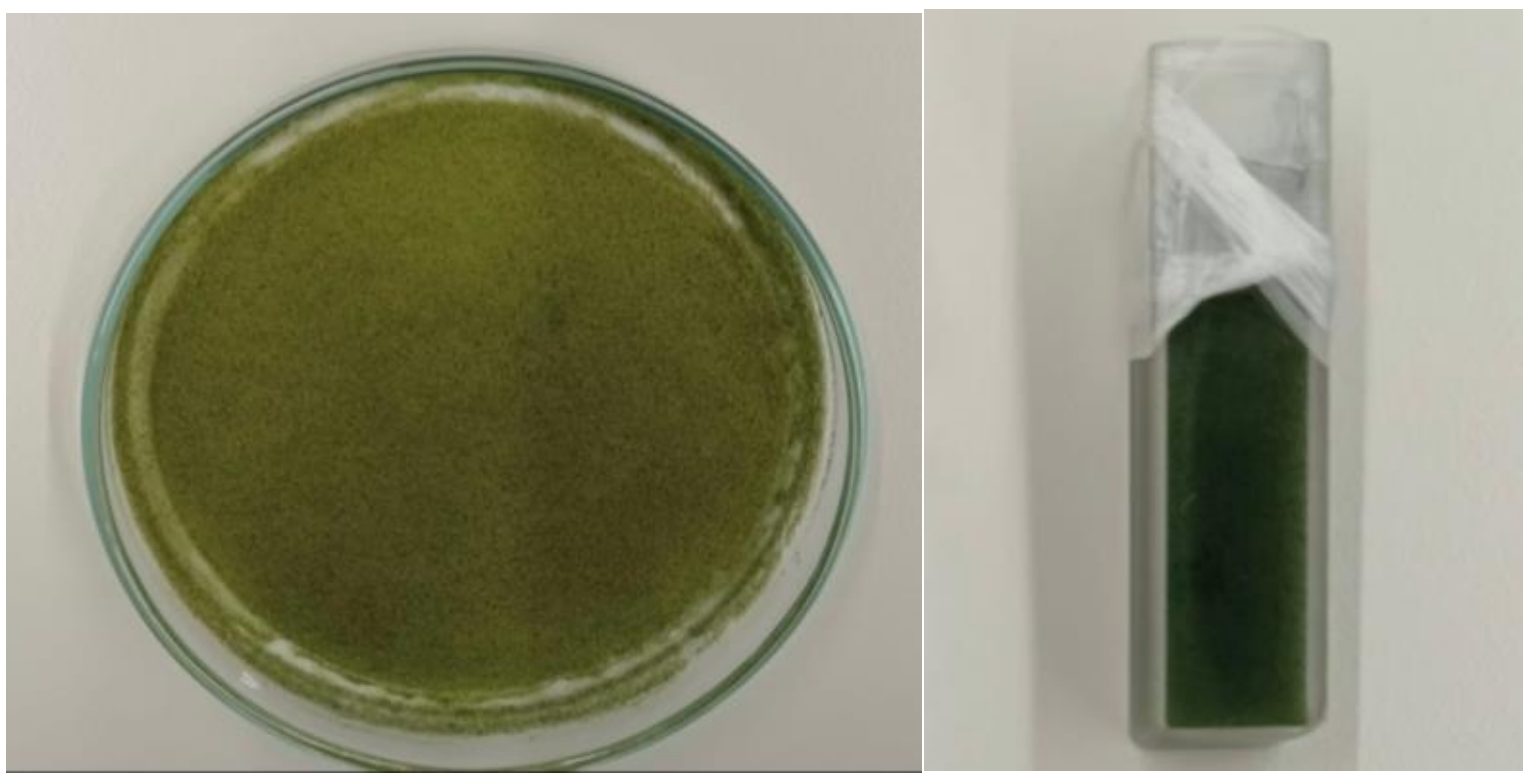

**Figure 3.** Sample 4 (0.06g *Chlorella*, 0.24g *k-Carrageenan*, 0.3g NaCl, distilled 5mL water) that obtained the best gel formation properties inside of a Petri dish and a quartz cuvette.

Film formation of the best gel form sample in *Chlorella- k-Carrageenan* gel formation (Sample 4) was started by preparing another sample that includes 0.06g *Chlorella*, 0.24g *k-Carrageenan*, 0.3g NaCl, 5mL water into a different 5mL beaker. The sample was heated on a heating block and mixed with a magnetic stirrer bar that was located inside of the beaker from 20 to 80 C° around 1 hour at room temperature (20 C°). When the temperature reached the range of 40-50 C°, the sample was poured into a square mold that had a 5cm length 5cm width, and 1.8cm depth for creating film that was required for photometric analyses.

The best samples in each gel formation trial group were chosen to obtain better analyses of observing gel formation. For this, heating, and cooling by going into sol-gel and gel-sol phases were performed in a spectrophotometer.

In the heating step that belonged to the gel-sol phase, a water bath enabled hot water to spread around the sample that was located inside the quartz cuvette in the spectrophotometer. After filling the water bath, pipes that contained copper parts were placed into another 1L water bath on the heating block to provide heat to the sample. To follow changes in the temperature, a digital multimeter was put inside the quartz cuvette. Lastly, power supplies that operated a 12V motor source to transfer water all around the pipes were turned on with the heating block that heated the 1L water bath. Significantly, the motor source was settled in the voltage range of 4-5V for constant water circulation and proper data collection. Sample collection by performing a video recording on a tripod for recording temperature and absorbance values also took place. Before starting the video recording, the reference value was recorded by locating a quartz cuvette that contained only water into the photometer at 550nm and 640nm separately. When the reference value was recorded at 550nm, the heating process started right after at 550nm at 20 C° by covering the top of the spectrophotometer with a piece of cardboard to prevent any light passages. When the thermometer reached 80 C°, the heating block was turned off and preparations for cooling processes were started.

In the cooling part of the photometric analyses which promoted sol-gel formation, another video recording of data collection was started directly when the heating block was turned off and a 1L water bath that contained pipes that contained copper parts was taken away from the heating block. To speed up the cooling, cold water is added in small amounts inside the water bath to cool the circulating water easily. When the temperature dropped to 20 C°, the power supply that operated the motor source was turned off to start the heating and cooling procedure at 640nm. At 640 nm, the same heating and cooling procedures that were used for 550 nm were redone.

**5.2 Biodegradability Under Soil Tests:**

Since the best sample at gel formation was detected as 0.06g *Chlorella*, 0.24g *k-Carrageenan*, 0.3g NaCl, 5ml water, to check its biodegradability five different samples weighing 0.28955g, 0.23775g, 0.25.354g, 0.25042g, and 1.12482g were prepared by using an analytic balance (Figure 4) (Figure 5). Samples were numbered 1 to 5 for a clear track. After the preparation step, samples were buried in the middle part of a 25mL beaker that was filled with soil. Exceptionally, sample 5 (1.12482g) was buried into a 50 mL beaker. For data collection, samples were checked by taking them out of the soil, removing the excess soil by washing them with water, and weighing them using analytic balance before putting them back into the soil at the same level in daily and weekly periods.

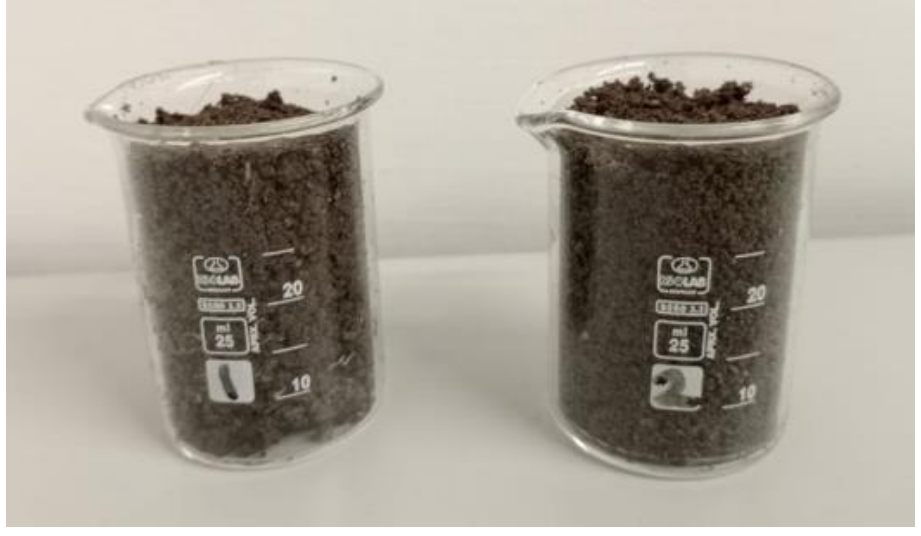

**Figure 4.** Sample 1-2 in 25mL beakers.

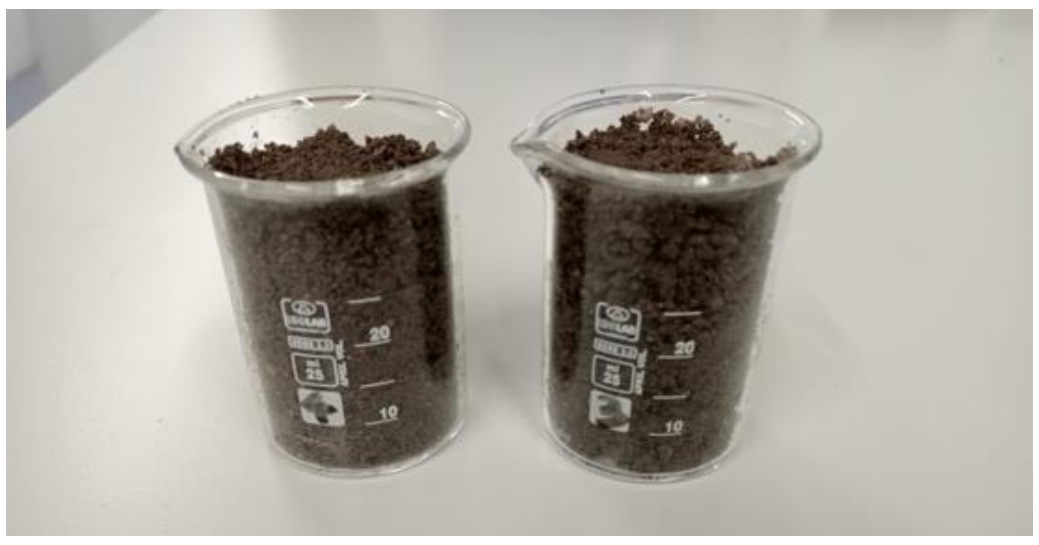

**Figure 5.** Sample 3-4 in 25mL beakers.

To support biodegradability and check how long the best sample (0.06g *Chlorella*, 0.24g *k-Carrageenan*, 0.3g NaCl, 5mL distilled water) did not leak any water without using protector films, a cup-shaped sample was prepared by weighing dry components in an analytic balance, adding distilled water, heating, and mixing the sample and pouring into a 50 mL beaker with a 25mL beaker inside of it. After waiting a few minutes to make sure the cup shape was properly formed, the sample was taken out of the beaker, transferred onto a plate, and poured water inside to observe water resistance (Figure 6). Data collection was done by recording the time until the water leaked out of the cup.

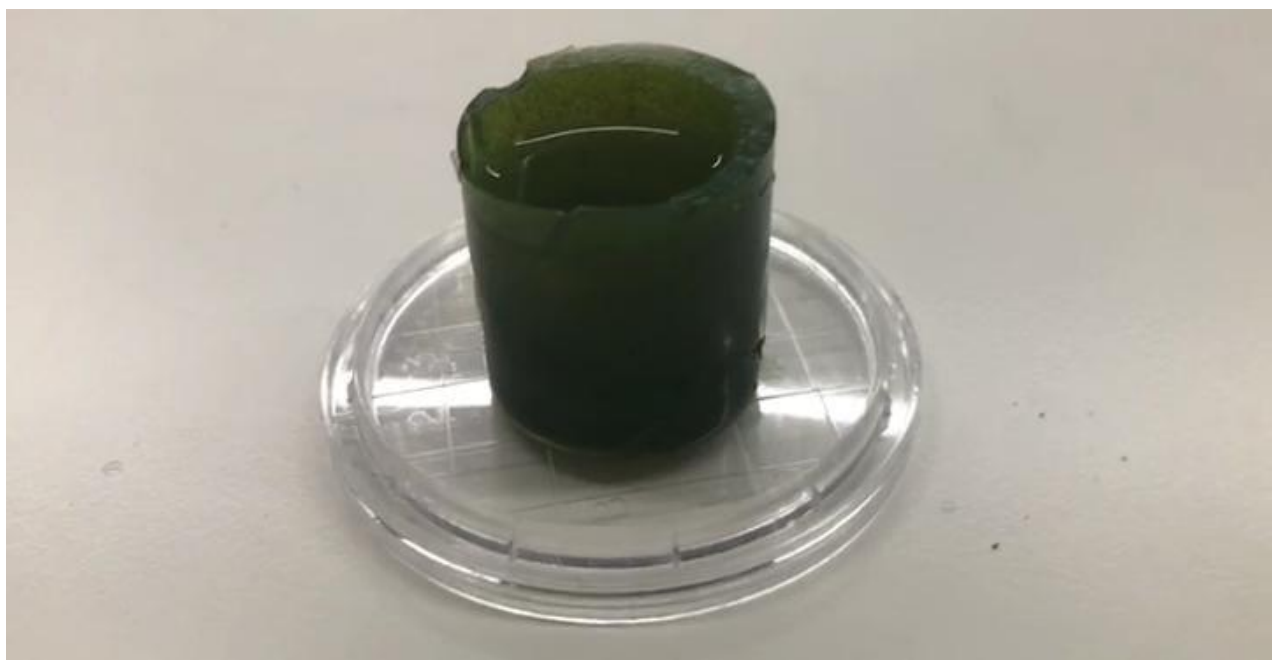

**Figure 6.** The best sample that had complete gel formation was in the water resistance test.

### 5.3 Photon Transmission Analyses

The analysis was conducted under 550 and 640 nanometers (nm), heating and cooling were performed. From the collected data, absorbance, and transmittance (TM) plots were obtained. For Sample 1, the following absorbance and transmittance were obtained (Figure 7).

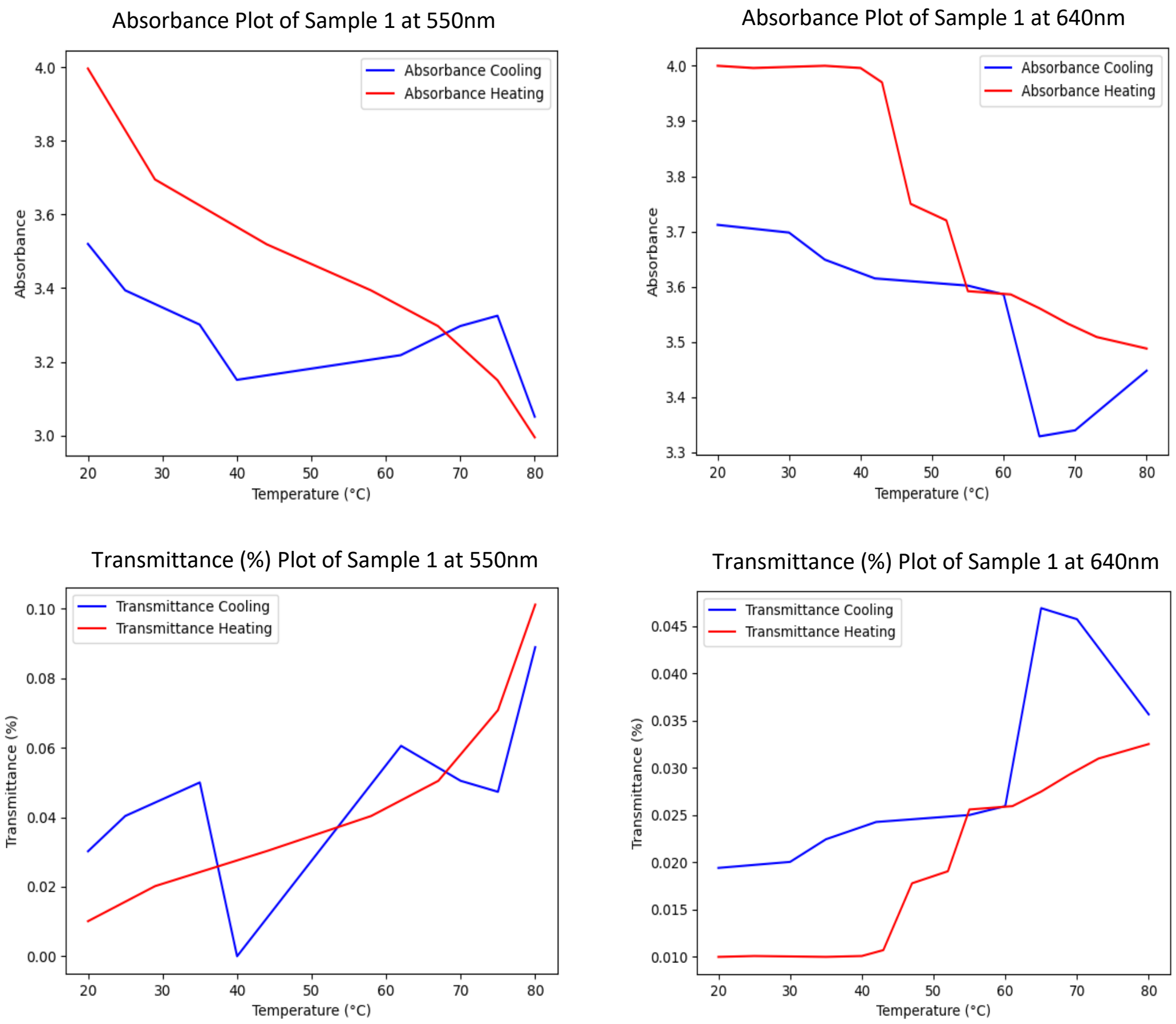

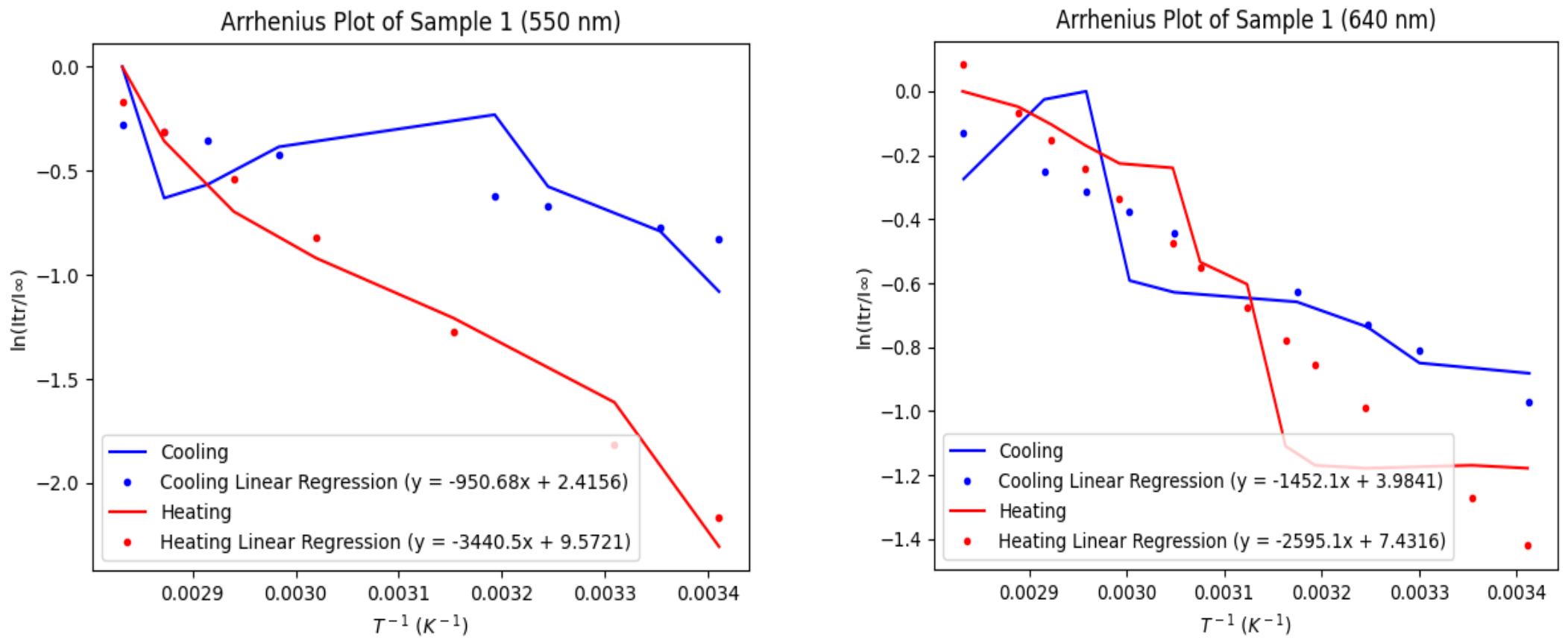

**Figure 7.** The absorbance, transmittance, and Arrhenius plot measurements were done under 550 and 640 nm for Sample 1 (0.06g *Chlorella* 0.06g *k-Carrageenan* 0.3g NaCl, 5 mL distilled water).

For Sample 2, heating, and cooling under 550 and 640 nm were repeated to obtain the absorbance, transmittance, and Arrhenius plots (Figure 8).

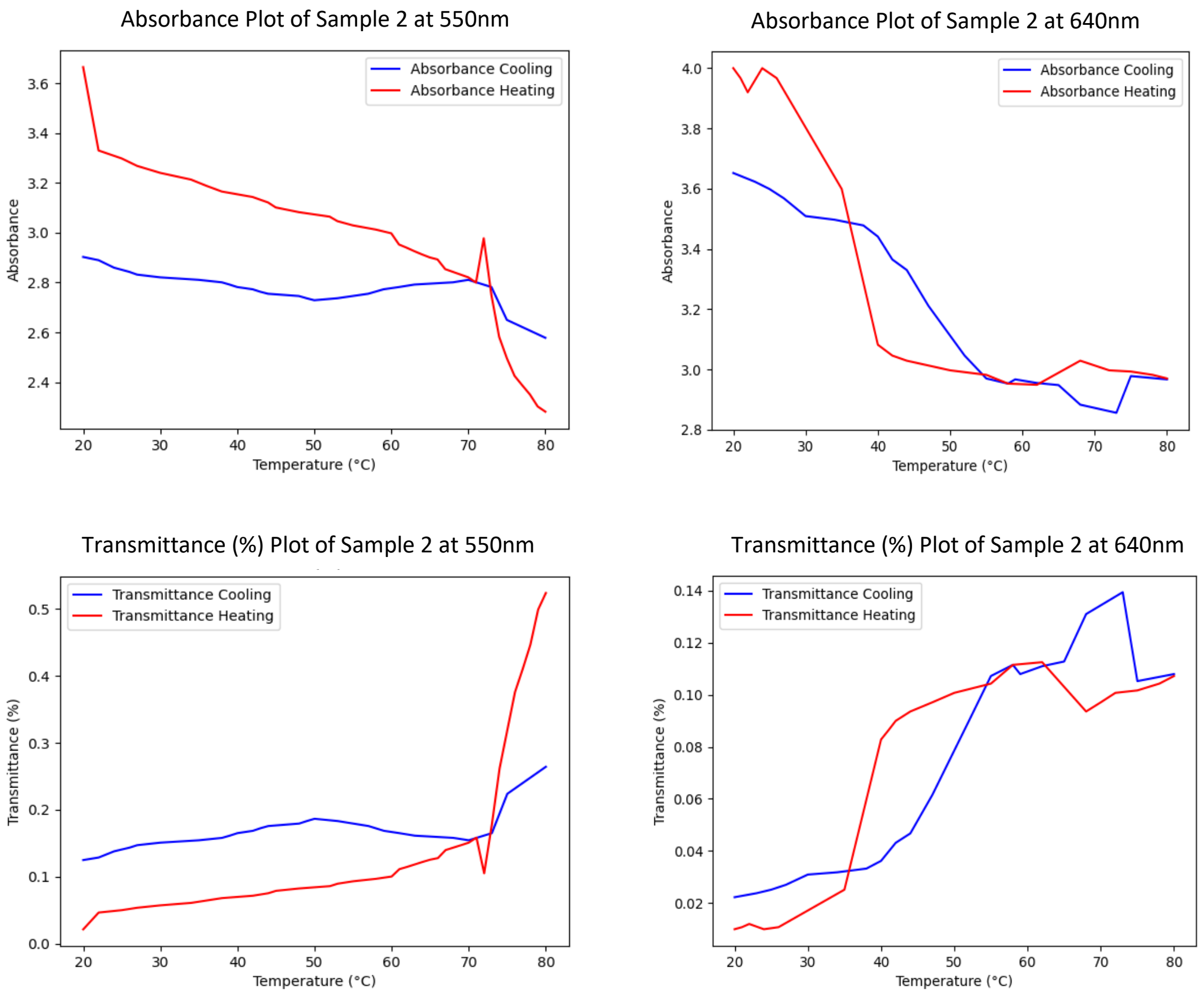

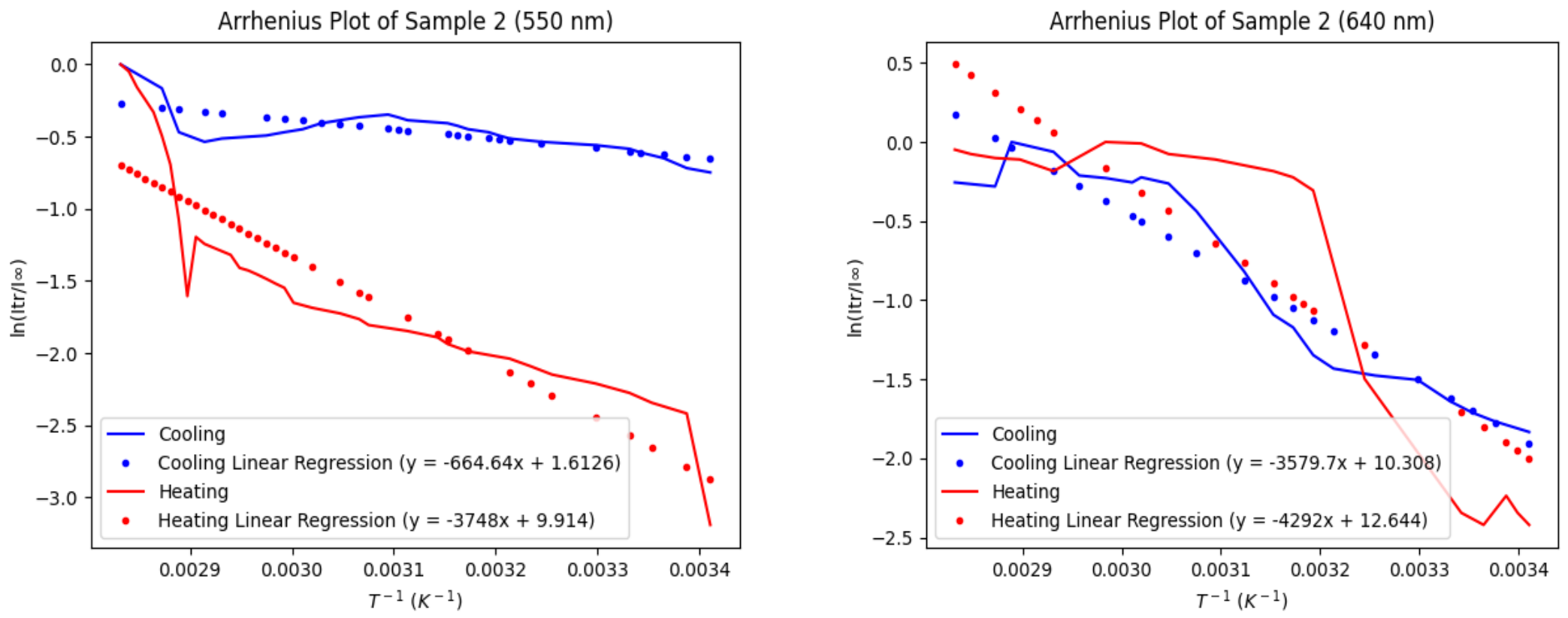

**Figure 8.** The absorbance, transmittance, and Arrhenius plot measurements were done under 550 and 640 nm for Sample 2 (0.06g *Chlorella*, 0.12g *k-Carrageenan*, 0.3g NaCl, distilled 5mL water).

The absorbance, transmittance, and Arrhenius plots of the heating-cooling for Sample 3 came out as following (Figure 9).

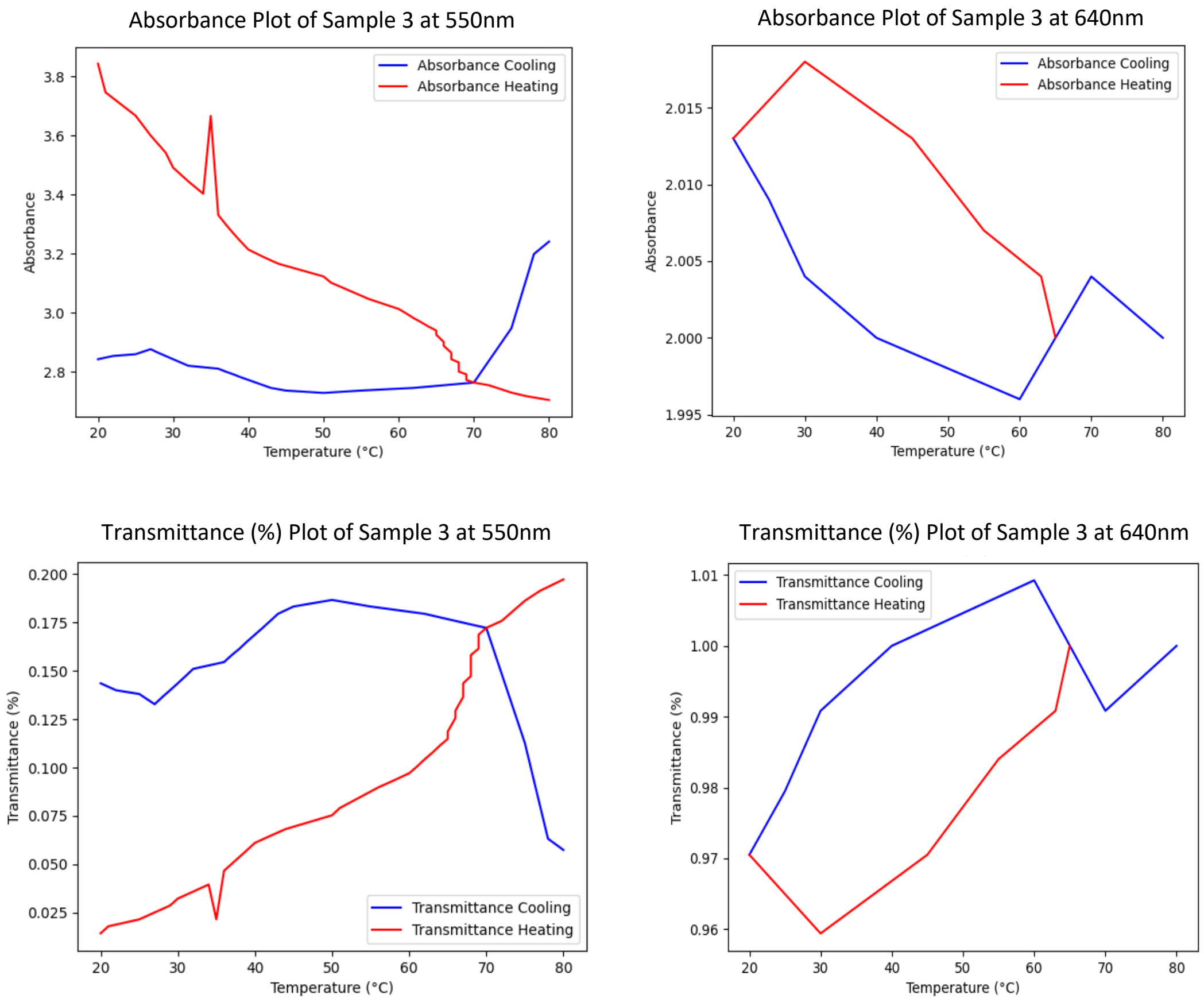

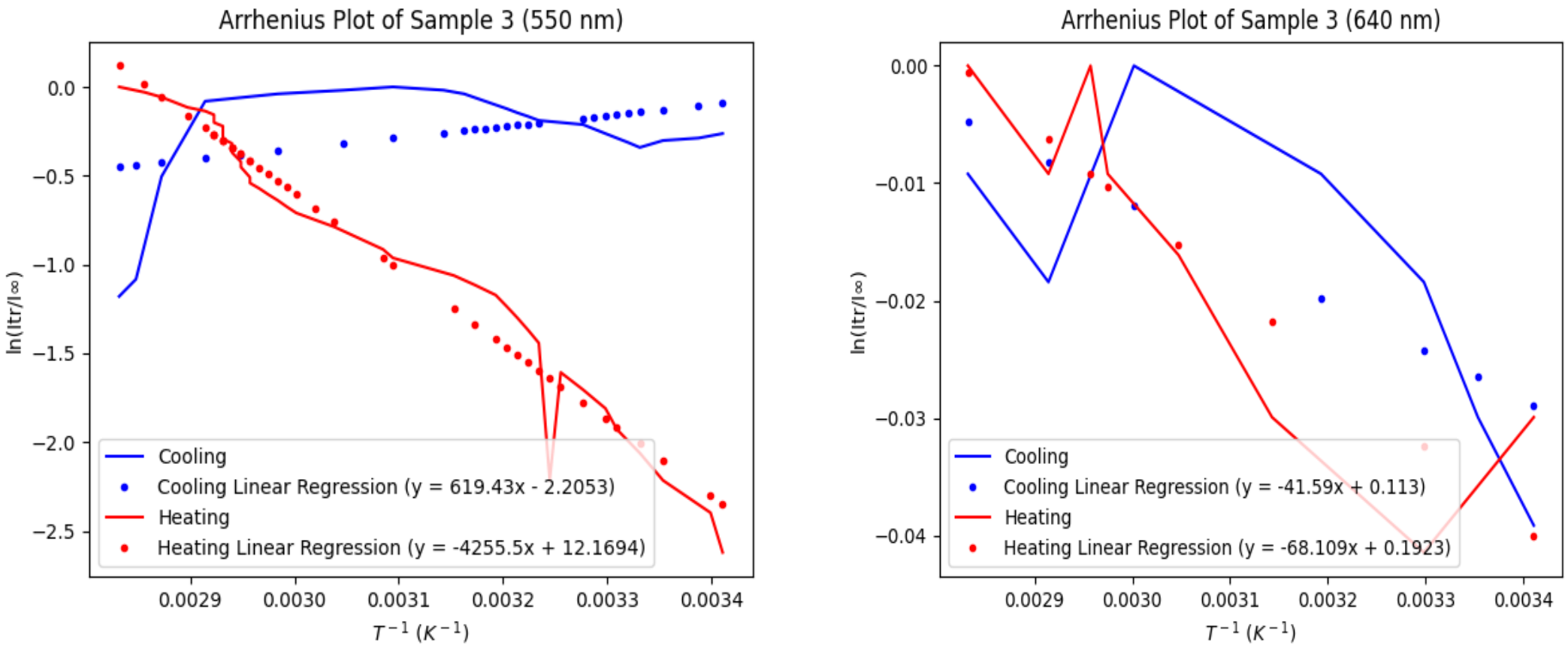

**Figure 9.** The absorbance, transmittance, and Arrhenius plot measurements were done under 550 and 640 nm for Sample 3 (0.06g *Chlorella*, 0.18g *k-Carrageenan*, 0.3g NaCl, distilled 5mL water).

Sample 4's absorbance, transmittance, and Arrhenius plots were drawn as following (Figure 10).

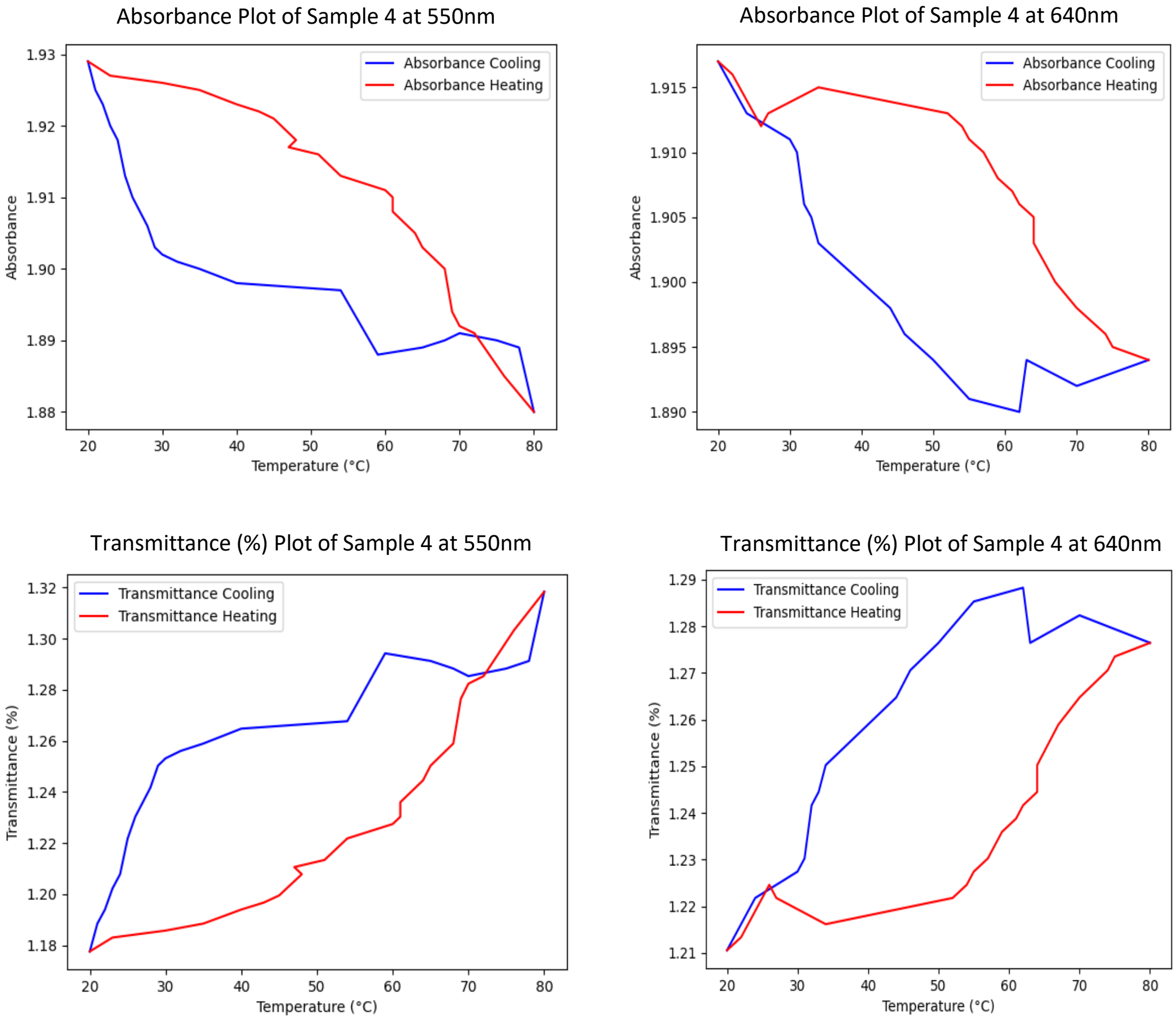

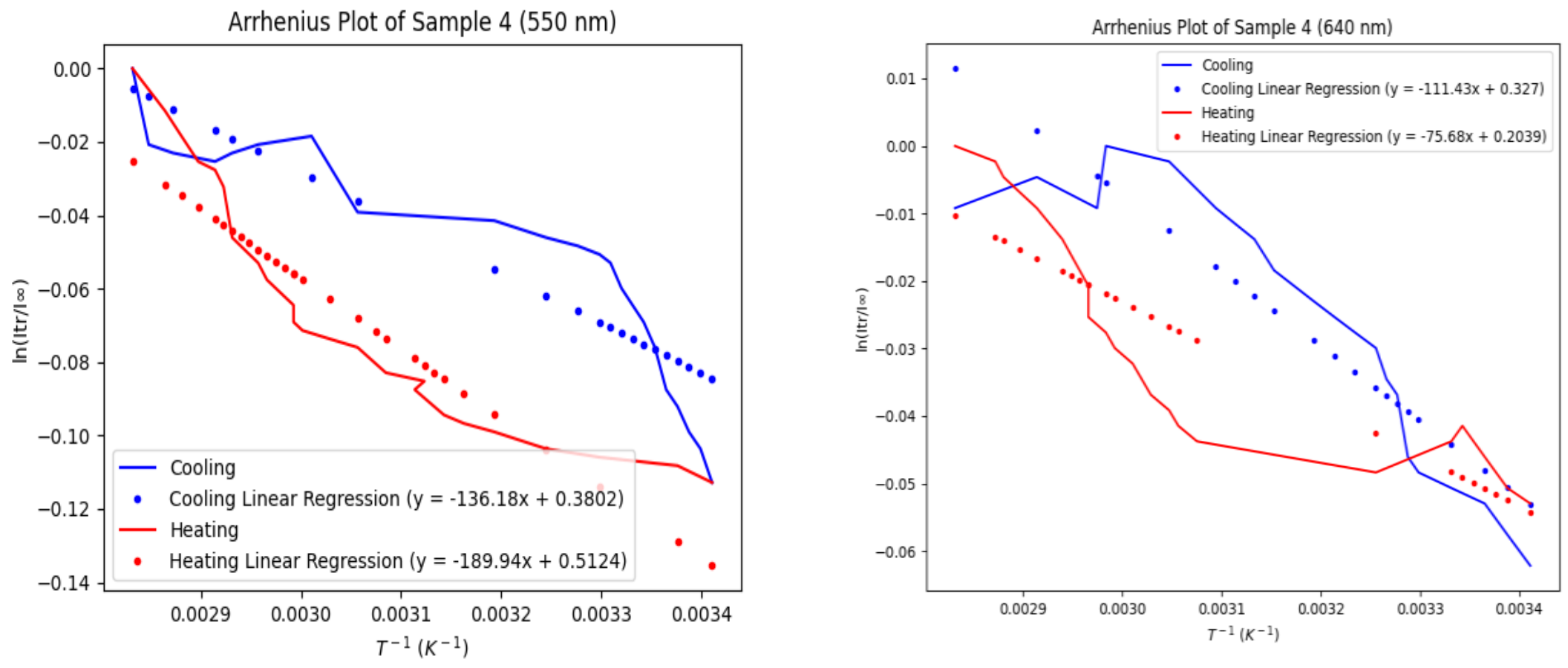

**Figure 10.** The absorbance, transmittance, and Arrhenius plot measurements were done under 550 and 640 nm for Sample 4 (0.06g *Chlorella*, 0.24g *k-Carrageenan*, 0.3g NaCl, distilled 5mL water).

Sample 5's absorbance, transmittance, and Arrhenius plots are provided below (Figure 11).

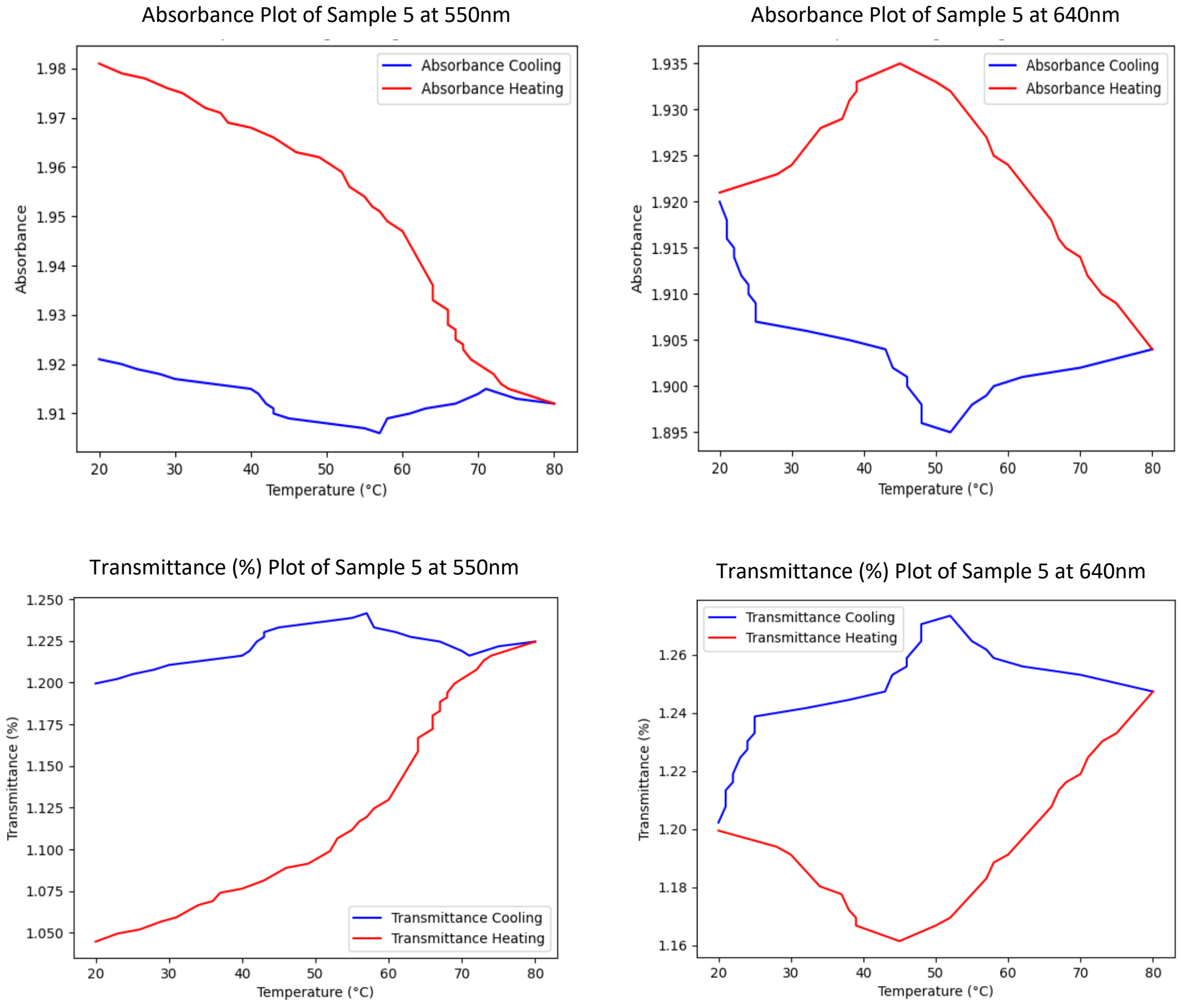

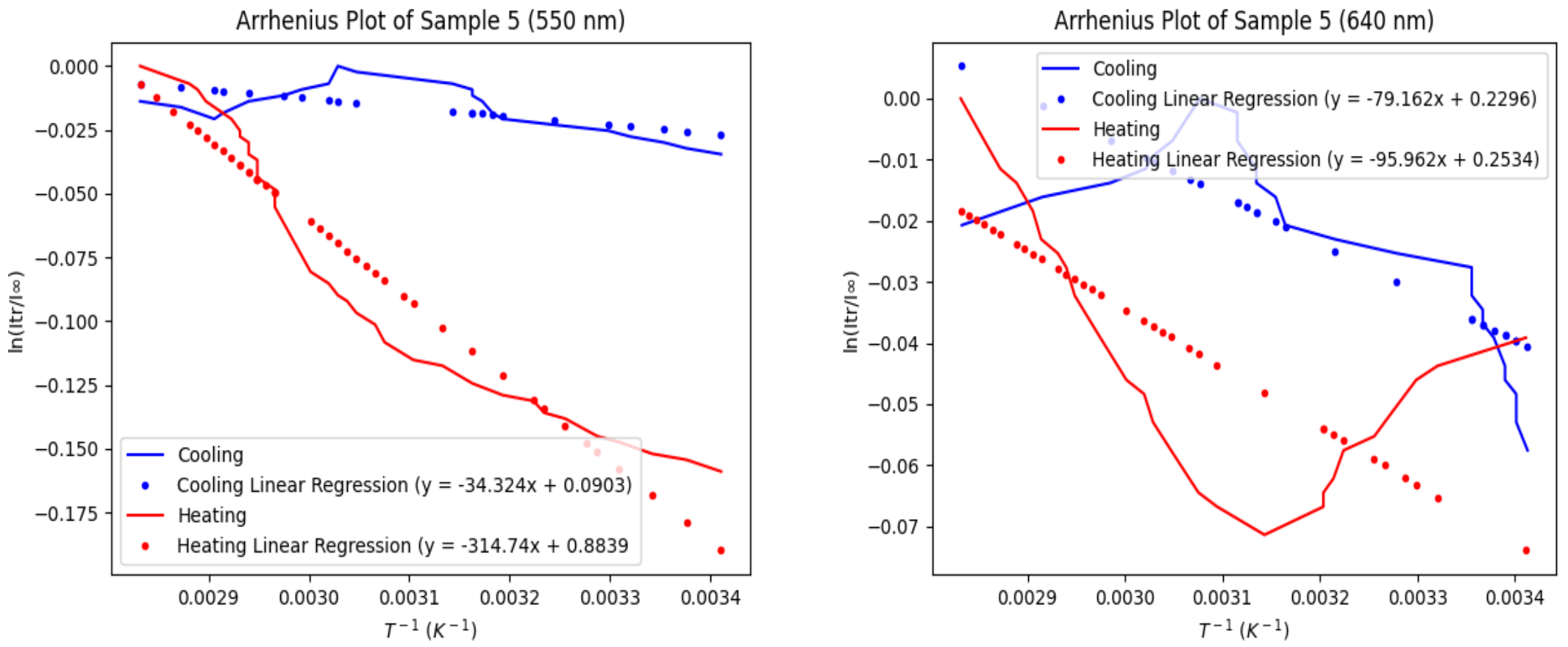

**Figure 11.** The absorbance, transmittance, and Arrhenius plot measurements were done under 550 and 640 nm for Sample 5 (0.06g *Chlorella*, 0.30g *k-Carrageenan*, 0.3g NaCl, distilled 5mL water).

By using slope values of the Arrhenius plots, activation energies were calculated (Table 3). Produced Energy values for Gel-Sol and Sol-Gel transition are given in Table 4.

**Table 3.** Slope values of Arrhenius plots of each sample.

| | 550 nm | | 640 nm | |
|---|---|---|---|---|
| **Sample #** | Slope Heating | Slope Cooling | Slope Heating | Slope Cooling |
| **Sample 1** | -3440.5 | -950.68 | -2595.1 | -1452.1 |
| **Sample 2** | -3748 | -664.64 | -4292 | -3579.7 |
| **Sample 3** | -68.109 | -41.59 | -4255.5 | 619.43 |
| **Sample 4** | -189.94 | -136.18 | -75.68 | -111.43 |
| **Sample 5** | -314.74 | -34.324 | -95.962 | -79.162 |

**Table 4.** Gel–Sol (heating) and Sol–Gel (cooling) activation energies (ΔEgs and ΔEsg).

| | 550 nm | | 640 nm | |
|---|---|---|---|---|
| **Sample #** | ΔEgs (kJ/mol) | ΔEsg (kJ/mol) | ΔEgs (kJ/mol) | ΔEsg (kJ/mol) |
| **Sample 1** | 2.48662771 | 3.34117545 | 2.633992967 | 0.00001031 |
| **Sample 2** | 20.7242034 | 9.16589605 | 88.3932575 | 53.29057 |
| **Sample 3** | 73.5757151 | 13.5639549 | 0.84606119 | 0.41294502 |
| **Sample 4** | 3.67582073 | 4.90318163 | 1.79897061 | 1.3517499 |
| **Sample 5** | 4.82694212 | 0.820949302 | 2.21177342 | 8.83599646 |

In addition to absorbance, transmittance, and Arrhenius plots, since Sample 4 was deemed best at gel formation, infrared spectroscopy was also performed on the sample to get a better understanding of the types of bonds present in the sample. After the spectroscopy, the transmittance and the wavelength were plotted against each other (Figure 12) (Figure 13).

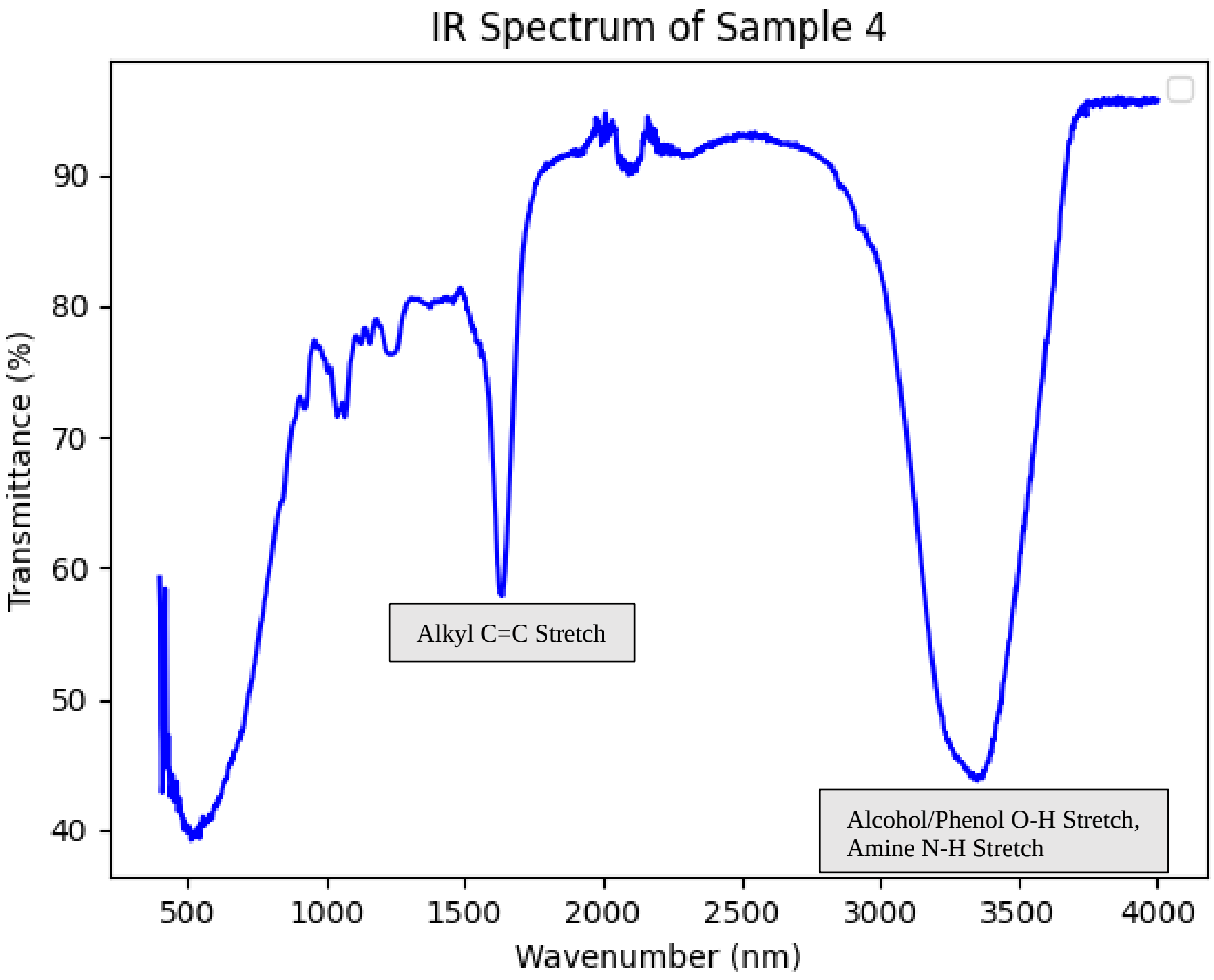

**Figure 12.** The IR Spectrum graph of Sample 4.

**IR Absorptions of Common Functional Groups**

| *Functional Group* | *Absorption Location ($cm^{-1}$)* | *Absorption Intensity* |
|---|---|---|
| Alkane (C–H) | 2,850–2,975 | Medium to strong |
| Alcohol (O–H) | 3,400–3,700 | Strong, broad |
| Alkene (C=C)<br>(C=C–H) | 1,640–1,680<br>3,020–3,100 | Weak to medium<br>Medium |
| Alkyne<br>(C≡C)<br>(C≡C–H) | <br>2,100–2,250<br>3,300 | <br>Medium<br>Strong |
| Nitrile (C≡N) | 2,200–2,250 | Medium |
| Aromatics | 1,650–2,000 | Weak |
| Amines (N–H) | 3,300–3,350 | Medium |
| Carbonyls (C=O)<br>Aldehyde (CHO)<br>Ketone (RCOR)<br>Ester (RCOOR)<br>Acid (RCOOH) | <br>1,720–1,740<br>1,715<br>1,735–1,750<br>1,700–1,725 | Strong |

**Figure 13.** Functional groups in the IR spectrum table [23].

Furthermore, to get a better understanding on was the Sample 4 photodegradable, photodegradation analyses under UV-A, UV-B, and UV-C lights were performed for 3 hours. When the sample showed the same result in all UV lights (Figure 14), another modified plot (Figure 15) was done by performing changes in absorbance values.

Photodegradation Results of Sample 4

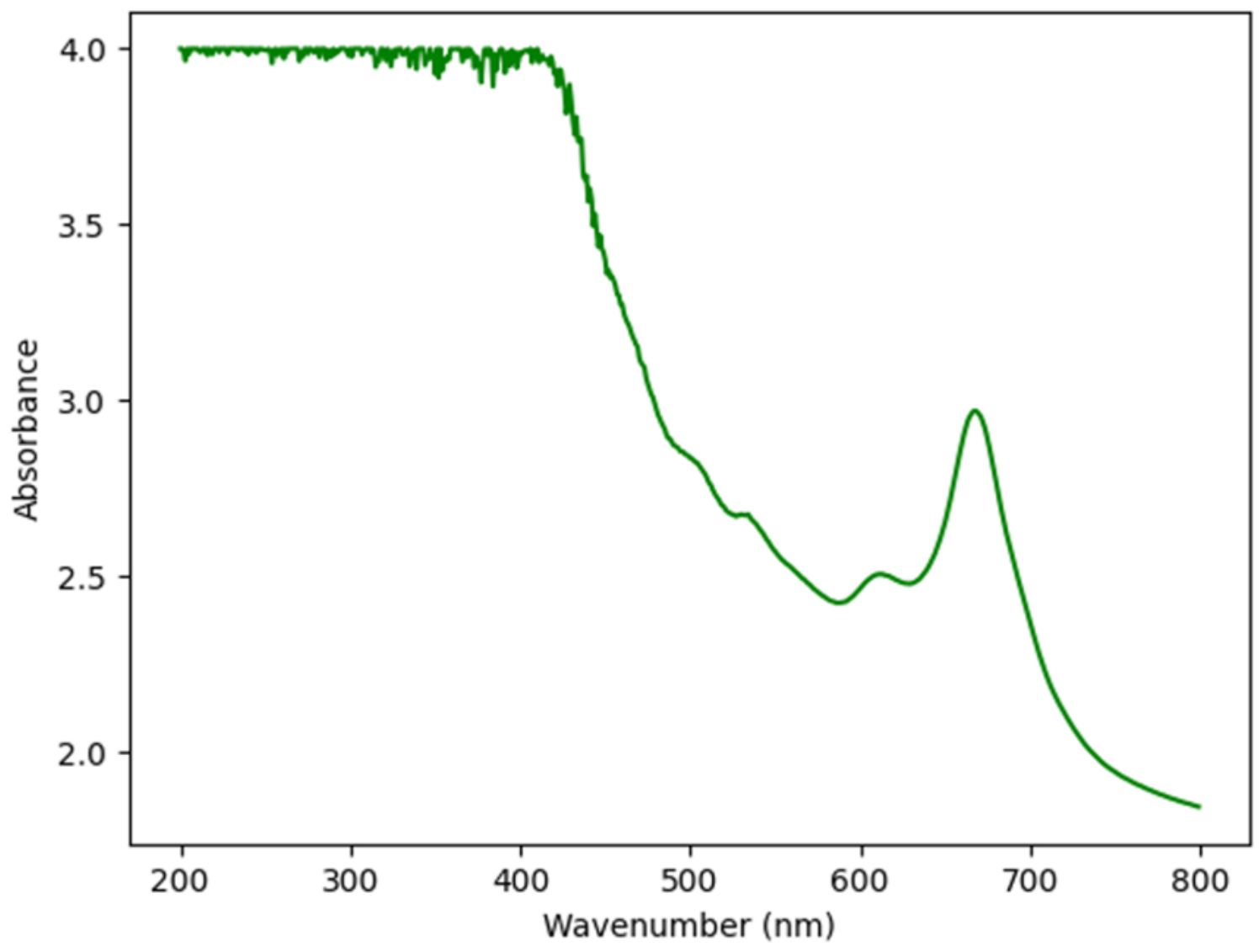

**Figure 14.** Photodegradation outcomes of Sample 4. Since all the lines overlapped, only the green color appeared.

Detailed Photodegradation Results of Sample 4

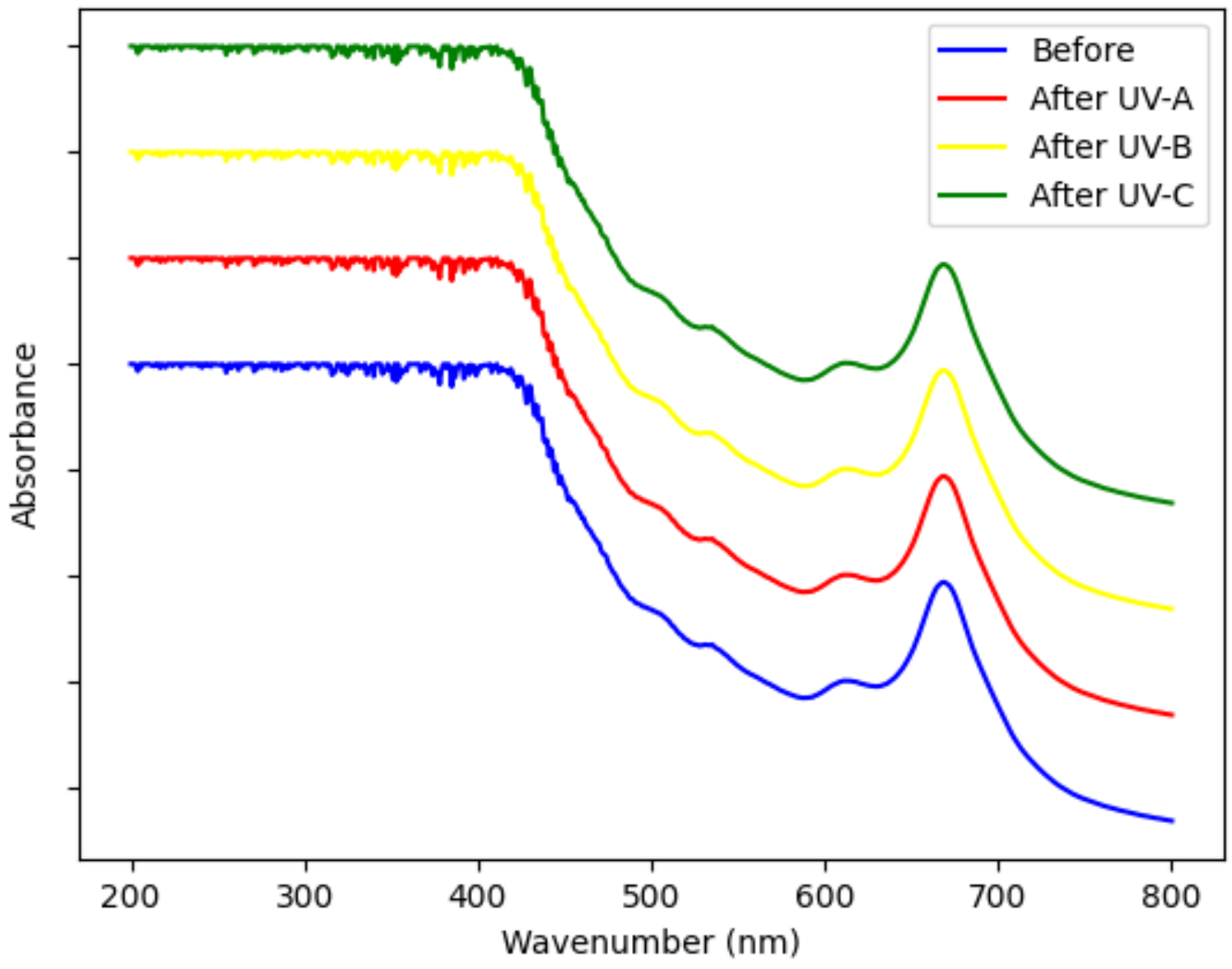

**Figure 15.** Detailed representation of photodegradation results of Sample 4. Section ‘Before’ represents the first outcome that did not interact with any UV light.

### 5.4 Biodegradability Results

All samples in this part belonged to sample 4 (0.06g *Chlorella*, 0.24g *k- Carrageenan*, 0.3g NaCl, 5 mL water).

**Sample 1 Physical Features:**

0.5cm width, 0.93cm length, 0.42cm depth

**Soil Amount in Sample 1:** 48.13875g

**Sample 2 Physical Features:**

0.5cm width, 0.90cm length, 0.44cm depth

**Soil Amount in Sample 2:** 39.52681g

**Table 5.** Biodegradability rates of samples 1 and 2 weekly.

| Dates | Sample 1 | Sample 2 |
|---|---|---|
| **18.04.2023** | 0.28955g | 0.23775g |
| **24.04.2023** | 0.03683g | 0.0g |
| **25.04.2023** | 0.0g | 0.0g |

**Sample 3 Physical Features:**
0.55cm width, 0.90cm length, 0.40cm depth
**Soil Amount in Sample 3:** 42.15196g

**Sample 4 Physical Features:**
0.60cm width, 0.50cm length, 0.45cm depth
**Soil Amount in Sample 4:** 42.5403g

**Table 6.** Biodegradability rates of samples 3 and 4 daily.

| Dates | Sample 3 | Sample 4 |
|---|---|---|
| **24.04.2023** | 0.25354g | 0.25042g |
| **25.04.2023** | 0.08073g | 0.05403g |
| **26.04.2023** | 0.07893g | 0.016801g |
| **27.04.2023** | 0.05684g | 0.0g |
| **28.04.2023** | 0.04174g | 0.0g |
| **02.05.2023** | 0.0g | 0.0g |

**Sample 5 Physical Features:**

2cm width, 1.5cm length, 1.5cm depth

**Soil Amount in Sample 5:** 64.16174g

**Table 7.** Biodegradability rates of sample 5 on a weekly and daily basis.

| Dates | Sample 5 |
|---|---|
| **02.05.2023** | 1.12482g |
| **08.05.2023** | 0.29564g |
| **09.05.2023** | 0.22953g |
| **10.05.2023** | 0.19852g |
| **11.05.2023** | 0.0g |

The cup-shaped sample was able to show water resistance for 1 day (09.05.2023-10.05.2023) without containing protector films inside of the surface.

**6. DISCUSSION**

From the start of the experiment, it was observed from its physical form that *Chlorella* alone was not enough to form a gel, as the samples were too liquid. Even after the trials with acetone and glycerin, the proper physical gel form was not observed. Since glycerin is a lipid-based molecule [24] and *Chlorella* dissolves in water, this could be the reason for the failure [25]. As for the acetone, since it contains both polar and non-polar bonds [26], proper interaction with *Chlorella* might not have taken place. Another reason this could be because *Chlorella* was obtained commercially and was not pure. Because it boosts gel formation by promoting the formation of random chains and cross-linking of the chains, table salt (NaCl) was also added to these samples, but there still were no promising results [27].

At the end of gel formation experiments, all samples provided outstanding physical forms. However, only three of them provided sigmoidal curves in their plots. From these three samples, sample 3 did not have a sigmoidal curve at 550nm. Additionally, while performing well at other wavelengths, sample 5 did not have a sigmoidal curve at 550nm for the absorbance. Sample 4 provided sigmoidal curves at all wavelengths and had a better physical form therefore it was chosen as the best sample.

In addition, Arrhenius plots were provided starting from the trials that contained *k-Carrageenan*. Arrhenius plots enabled the calculation of negative activation energy of the sol-gel/gel-sol processes through the slopes [13]. The activation energies were informative about the reaction rates. Overall, the *k-Carrageenan* addition into the samples lowered the energy needs for both transitions during their hysteresis cycles. For the best sample (Sample 4), the Arrhenius plot at 640 nm was as expected, where the gel-sol transition had lower activation energy than the sol-gel transition. However, at 550 nm it was observed that gel-sol transition had higher activation energy than sol-gel transition. This exception might be how the material behaves, but it could also be caused because the *Chlorella* used in the experiment was obtained commercially. The exact reason for this exception is unknown, as it is only observed in Sample 4, and it needs further research. Since the results at 640 nm were in the expected range, the results at that wavelength were more reliable.

Through the IR spectra, information about the characteristics of different chemical bonds can be obtained. Samples can be characterized and chemically discriminated. So, to have further information about the bond types and photodegradability, an IR spectroscopy was performed on sample 4. Since IR spectrometry can be used to obtain information about different bond signals, some bonds present in Sample 4 were identified [28]. When the peaks on the IR plot were compared to the IR absorption table, the first three peaks (543.4782609, 1039.130435, 1249.689441) were in the fingerprint region (500-1450nm), therefore they were insignificant. As for the fourth peak (1637.267081), an Alkyl C=C Stretch appears to be present, when the peak was compared to the IR absorption table. Another insignificant peak appeared at 2093.78882. Finally, at 3359.006211 both an alcohol/phenol O-H stretch, and an amine N-H stretch were observed. When the UV spectroscopy plots obtained from Istanbul Technical University were observed, since all three lines are aligned at the plot, it can be said that sample 4 was not photodegradable. The sample not being photodegradable means it is durable, and in turn, makes it a good replacement material for plastic in the packaging [8].

After finishing the photometric analyses, the biodegradability analyses were done for sample 4 only due to it being the best sample. There were two types of biodegradability analyses: degradability in soil and water resistance. The first analysis was the soil degradability, and it was done in three parts. In the first part, samples with two different weights were buried in the soil and checked weekly. It was seen that they had dissolved completely within a week. To get more detailed information, the second part of the analysis was done where the samples were checked daily. While sample 3 dissolved in 8 days, sample 4 dissolved in only 4 days. This difference could be since the soil was obtained directly from nature and was not controlled. To check if a bigger sample could dissolve in soil, an amount weighing 1.12482g was buried and checked weekly and daily. Even with a great difference in weight, it was observed that sample 5 also dissolved quickly like the other samples. Judging by these results, the best sample (0.06g *Chlorella*, 0.24g *k-Carrageenan*, 0.3g NaCl, distilled 5mL water) is greatly biodegradable.

After the soil degradability tests were completed, the water resistance tests began. These tests were conducted to prove whether the material was suitable for packaging. The material was shaped into a small cup and filled with water. The material was able to hold the water for an entire day without leaking. It has to be noted that the material was not lined with any protector film, hence this test proved that the material was more than enough to be used in packaging.

## 7. CONCLUSIONS

Plastic is a huge environmental hazard, especially in the form of packaging. As an alternative, biodegradable packaging options can be considered. Microalgae, *Chlorella* specifically, was a great option for its favorable characteristics for packaging. After performing several experiments, it was decided to combine *Chlorella* with *k-Carrageenan* to obtain an optimum packaging material. Upon the combination of two different microalgae with NaCl and water, a proper gel formation was observed physically and photometrically. Considering the results of soil degradability tests, the material proved itself to be highly biodegradable. In addition to that, the water resistance test results, and the fact that it is not photodegradable proved that this material was suitable to use for packaging.

**REFERENCES:**

[1] M.I. Khan, J.H. Shin, J.D. Kim, The promising future of microalgae: current status, challenges, and optimization of a sustainable and renewable industry for biofuels, feed, and other products, Microb Cell Fact. 17 (2018) 36. https://doi.org/10.1186/s12934-018-0879-x.

[2] L. Li, R. Ni, Y. Shao, S. Mao, Carrageenan and its applications in drug delivery, Carbohydr Polym. 103 (2014) 1–11. https://doi.org/https://doi.org/10.1016/j.carbpol.2013.12.008.

[3] I.T.K. Ru, Y.Y. Sung, M. Jusoh, M.E.A. Wahid, T. Nagappan, Chlorella vulgaris: a perspective on its potential for combining high biomass with high value bioproducts, Applied Phycology. 1 (2020) 2–11. https://doi.org/10.1080/26388081.2020.1715256.

[4] W.G. Morais Junior, M. Gorgich, P.S. Corrêa, A.A. Martins, T.M. Mata, N.S. Caetano, Microalgae for biotechnological applications: Cultivation, harvesting and biomass processing, Aquaculture. 528 (2020) 735562. https://doi.org/https://doi.org/10.1016/j.aquaculture.2020.735562.

[5] M.C. Meghana, C. Nandhini, L. Benny, L. George, A. Varghese, A road map on synthetic strategies and applications of biodegradable polymers, Polymer Bulletin. (2022). https://doi.org/10.1007/s00289-022-04565-9.

[6] P.B. Hatzinger, J.W. Kelsey, Biodegradation of organic contaminants, in: M.J. Goss, M. Oliver (Eds.), Encyclopedia of Soils in the Environment (Second Edition), Academic Press, Oxford, 2023: pp. 547–557. https://doi.org/https://doi.org/10.1016/B978-0-12-822974-3.00140-3.

[7] F. Alshehrei, Biodegradation of Synthetic and Natural Plastic by Microorganisms, J Appl Environ Microbiol. 5 (2017) 8–19. https://doi.org/10.12691/jaem-5-1-2.

[8] C. Guo, H. Guo, Progress in the Degradability of Biodegradable Film Materials for Packaging, Membranes (Basel). 12 (2022). https://doi.org/10.3390/membranes12050500.

[9] A. Kumar, P. Gajanan, Ey, A review on the factors affecting the photocatalytic degradation of hazardous materials, in: 2017. https://api.semanticscholar.org/CorpusID:44228134.

[10] S. Pfendler, B. Alaoui-Sossé, L. Alaoui-Sossé, F. Bousta, L. Aleya, Effects of UV-C radiation on Chlorella vulgaris, a biofilm-forming alga, J Appl Phycol. 30 (2018) 1607–1616. https://doi.org/10.1007/s10811-017-1380-3.

[11] E. Yousif, R. Haddad, Photodegradation and photostabilization of polymers, especially polystyrene: review, Springerplus. 2 (2013) 398. https://doi.org/10.1186/2193-1801-2-398.

[12] A. Nazir, A. Asghar, A. Aslam Maan, Chapter 13 - Food Gels: Gelling Process and New Applications, in: J. Ahmed, P. Ptaszek, S. Basu (Eds.), Advances in Food Rheology and Its Applications, Woodhead Publishing, 2017: pp. 335–353. https://doi.org/https://doi.org/10.1016/B978-0-08-100431-9.00013-9.

[13] S. Kara, E. Arda, B. Kavzak, Ö. Pekcan, Phase transitions of κ-carrageenan gels in various types of salts, J Appl Polym Sci. 102 (2006) 3008–3016. https://doi.org/https://doi.org/10.1002/app.24662.

[14] S. Banerjee, S. Bhattacharya, Food Gels: Gelling Process and New Applications, Crit Rev Food Sci Nutr. 52 (2012) 334–346. https://doi.org/10.1080/10408398.2010.500234.

[15] A.H. Karoyo, L.D. Wilson, Physicochemical Properties and the Gelation Process of Supramolecular Hydrogels: A Review, Gels. 3 (2017). https://doi.org/10.3390/gels3010001.

[16] L.A. Felton, Mechanisms of polymeric film formation, Int J Pharm. 457 (2013) 423–427. https://doi.org/https://doi.org/10.1016/j.ijpharm.2012.12.027.

[17] B. Lippold, R. Pagés, Film formation, reproducibility of production and curing with respect to release stability of functional coatings from aqueous polymer dispersions, Pharmazie. 56 (2001) 5–17.

[18] C. Burgess, Chapter 1 - The Basis for Good Spectrophotometric UV–Visible Measurements, in: O. Thomas, C. Burgess (Eds.), UV-Visible Spectrophotometry of Water and Wastewater (Second Edition), Elsevier, 2017: pp. 1–35. https://doi.org/https://doi.org/10.1016/B978-0-444-63897-7.00001-9.

[19] A. Asadova, E.A. Masimov, A.R. Imamaliyev, A.H. Asadova, Spectrophotometric investigation of gel formation in water solution of

agar, Modern Physics Letters B. 34 (2020) 2050147. https://doi.org/10.1142/S021798492050147X.

[20] M. Calcabrini, D. Onna, Exploring the Gel State: Optical Determination of Gelation Times and Transport Properties of Gels with an Inexpensive 3D-Printed Spectrophotometer, J Chem Educ. 96 (2019) 116–123. https://doi.org/10.1021/acs.jchemed.8b00529.

[21] M. Dadi, M. Yasir, Spectroscopy and Spectrophotometry: Principles and Applications for Colorimetric and Related Other Analysis, in: A.K. Samanta (Ed.), Colorimetry, IntechOpen, Rijeka, 2022: p. Ch. 4. https://doi.org/10.5772/intechopen.101106.

[22] S.K. Yatirajula, A. Shrivastava, V.K. Saxena, J. Kodavaty, Flow behavior analysis of Chlorella Vulgaris microalgal biomass, Heliyon. 5 (2019). https://doi.org/10.1016/j.heliyon.2019.e01845.

[23] J.T. Moore, Chemistry for dummies, 2002. https://ci.nii.ac.jp/ncid/BB15616443.

[24] W. Dowhan, Synthesis and Structure of Glycerolipids, in: R.A. Bradshaw, P.D. Stahl (Eds.), Encyclopedia of Cell Biology, Academic Press, Waltham, 2016: pp. 160–172. https://doi.org/https://doi.org/10.1016/B978-0-12-394447-4.10020-3.

[25] L. Grossmann, S. Ebert, J. Hinrichs, J. Weiss, Formation and Stability of Emulsions Prepared with a Water-Soluble Extract from the Microalga Chlorella protothecoides, J Agric Food Chem. 67 (2019) 6551–6558. https://doi.org/10.1021/acs.jafc.8b05337.

[26] T.C. Chan, C.H.C. Chan, W.Y. Tang, N.W. Chang, Effects of Hydrogen Bonding on Diffusion of Aromatic Compounds in Acetone: An Experimental Investigation from 268.2 to 328.2 K, J Phys Chem B. 122 (2018) 9236–9249. https://doi.org/10.1021/acs.jpcb.8b08539.

[27] M.J. Lewis, 5 - Solid rheology and texture, in: M.J. Lewis (Ed.), Physical Properties of Foods and Food Processing Systems, Woodhead Publishing, 1996: pp. 137–166. https://doi.org/https://doi.org/10.1533/9781845698423.137.

[28] U. Blazhko, V. Shapaval, V. Kovalev, A. Kohler, Comparison of augmentation and pre-processing for deep learning and chemometric classification of infrared spectra, Chemometrics and Intelligent

Laboratory Systems. 215 (2021) 104367. https://doi.org/https://doi.org/10.1016/j.chemolab.2021.104367.